\documentclass[
reprint, 
superscriptaddress,
 amsmath,amssymb,
prl,
]{revtex4-1}

\usepackage{graphicx}
\usepackage{dcolumn}
\usepackage{bm}
\usepackage{todonotes}

\usepackage{color}

\newcommand{\pdiff}[2]{\frac{\partial #1}{\partial #2}}

\newcommand{\new}{\nonumber\\}
\newcommand{\abs}[1]{\left|#1\right|}
\newcommand{\bx}{\bm{x}}

\newcommand{\bh}{\bm{h}}

\newcommand{\atan}{{\rm atan}}
\newcommand{\tri}{{\rm Tri}_{\varepsilon}}
\usepackage{amsmath}	
\begin{document}

\preprint{APS/123-Qed} \title{ 
Jamming with tunable roughness
}


\author{Harukuni Ikeda}
 \email{hikeda@g.ecc.u-tokyo.ac.jp}
\affiliation{
Graduate School of Arts and Sciences, The University of Tokyo 153-8902, Japan
}

\author{Carolina Brito}
\affiliation{
Instituto de F\'isica, UFRGS, 91501-970, Porto Alegre, Brazil}

\author{Matthieu Wyart}
\affiliation{
Institute of Physics, EPFL, CH-1015 Lausanne, Switzerland}

\author{Francesco Zamponi}
\affiliation{ Laboratoire de Physique de l'\'Ecole Normale Sup\'erieure,
Universit\'e PSL, CNRS, Sorbonne Universit\'e, Universit\'e de Paris,
75005 Paris, France }


\date{\today}
	     
\begin{abstract}  
We introduce a new model to study the effect of surface roughness on the
jamming transition. By performing numerical simulations, we show that
for a smooth surface, the jamming transition density and the contact
number at the transition point both increase upon increasing
asphericity, as for ellipsoids and spherocylinders. Conversely, for a
rough surface, both quantities decrease, in quantitative agreement with
the behavior of frictional particles. Furthermore, in the limit
corresponding to the Coulomb friction law, the model satisfies a
generalized isostaticity criterion proposed in previous studies. We
introduce a counting argument that justifies this criterion and
interprets it geometrically.  Finally, we propose a simple theory to
predict the contact number at finite friction from the knowledge of the
force distribution in the infinite friction limit.
\end{abstract}


\maketitle

\paragraph*{Introduction. --}

Upon compression, a granular material suddenly acquires a finite
mechanical pressure at a certain jamming transition density $\varphi_J$
at which constituent particles start to touch each
other~\cite{van2009,durian1995,zhang2009,bernal1960,donev2004,jaoshvili2010,bi2015,franz2016,franz2019multilayer}. One
of the most popular and simplest models of the jamming transition is a
system consisting of frictionless spherical particles interacting via a
purely repulsive potential. A notable property of the model is the
so-called isostaticity: the number of degrees of freedom equals the
number of constraints imposed by the contacts among constituent
particles. A simple counting argument predicts that when a frictionless
spherical system is isostatic, the contact number per particle is $z=2d$
in $d$ spatial dimensions. Experiments~\cite{bernal1960} and numerical
simulations~\cite{ohern2003,goodrich2012finite} show that the contact
number at $\varphi_J$ indeed satisfies $z_J=2d$. Remarkably, recent
numerical and theoretical progress unveiled that isostatic systems,
which encompass some classes of neural
networks~\cite{franz2016,franz2017universality,franz2019multilayer,franz2019linear}
in addition to frictionless spherical particles, belong to the same
universality
class~\cite{wyart2005rigidity,wyart2005compress,degiuli2014force,charbonneau2014fractal,charbonneau2015buckling}.

However, in experiments, friction has a significant effect on the
jamming transition. Systematic numerical studies have been performed for
spherical particles with the famous Mohr-Coulomb law: the tangential
force $f_t$ between two particles in contact is proportional to the
displacement from the point of contact as long as $\abs{f_t} \leq \mu
f_n$, where $f_n$ denotes the normal force, and $\mu$ denotes the
friction coefficient~\cite{cundall1979}. When the tangential force
reaches the Coulomb threshold $\abs{f_t} = \mu f_n$, the contact breaks
and the particles start to slip with respect to each other. If we assume
that each contact constraints one translational motion and $d-1$
tangential motions, the counting argument predicts $z_J = d+1$ when the
system is isostatic~\cite{van2009,ewdards1999}.  However, numerical
simulations show that $z_J$ smoothly decreases from $2d$ upon increasing
$\mu$, and converges to $d+1$ only in the large friction limit
$\mu\to\infty$~\cite{unger2005,silbert2010jamming}.  Isostaticity thus
seems to be broken for frictional particles.  However, more recently, it
has been realized that more careful considerations are necessary to derive
the isostatic condition for frictional
particles~\cite{bouchaud2002granular,shundyak2007,henkes2010}. The point
is that a finite fraction of the fully mobilized contacts satisfy the
Coulomb threshold $\abs{f_t} = \mu f_n$, and those contacts do not
constrain the tangential motion. This observation leads to a {\it
generalized} isostaticity condition $z_J = d+1+2n_m/d$, where $n_m$
denotes the number of fully mobilized contacts per
particle~\cite{henkes2010}. Numerical simulations prove that frictional
particles indeed satisfy generalized isostaticity at $\varphi_J$ when slowly
equilibrated~\cite{shundyak2007,henkes2010}.

Compared to frictionless particles, studies of the jamming of frictional
particles, {\it e.g.}, to unveil the mechanisms yielding the generalized
isostaticity condition and their universality class, are still in their
infancy. A reason is the strong non-analiticity of the Coulomb law at
the Coulomb threshold $\abs{f_t}=\mu f_n$, which makes the contact
network ill-defined~\cite{henkes2010}, and the lack of a well-defined
potential energy~\cite{chattoraj2019}. A way to avoid this difficulty is
to revisit the microscopic origin of the empirical Coulomb friction
law. Although there are several possible origins of
friction~\cite{persson2013sliding}, here we focus on the geometric
friction caused by surface roughness, which has gained a lot of
attention due to the recent development of experimental techniques such
as 3D printing~\cite{athanassiadis2014particle,hsiao2019experimental},
and advanced computational techniques for complex-shaped
particles~\cite{alonso2008spheropolygons,papanikolaou2013}. In this
work, we construct a new model to take into account the effect of
surface roughness by means of a perturbative expansion around the
reference case of spherical disks. By performing numerical simulations,
we show that, for a smooth surface, $z_J$ of the model increases upon
increasing asphericity, suggesting that a small deviation from perfect
disks plays a similar role to the asphericity of convex-shaped
particles~\cite{donev2004,donev2007,williams2003,blouwolff2006,werf2018,brito2018,ikeda2019mean,ikeda2020}.
Contrarily, for a rough surface, $z_J$ decreases upon increasing
asphericity, as for frictional particles. Furthermore, we show that our
model gives a clear explanation for why particles with Coulomb friction
satisfy the generalized isostaticity condition.  Finally, we propose a
simple approximation scheme to calculate $z_J$ for frictional particles.

\paragraph*{Model. --}

\begin{figure} 
 \centering \includegraphics[width=6cm]{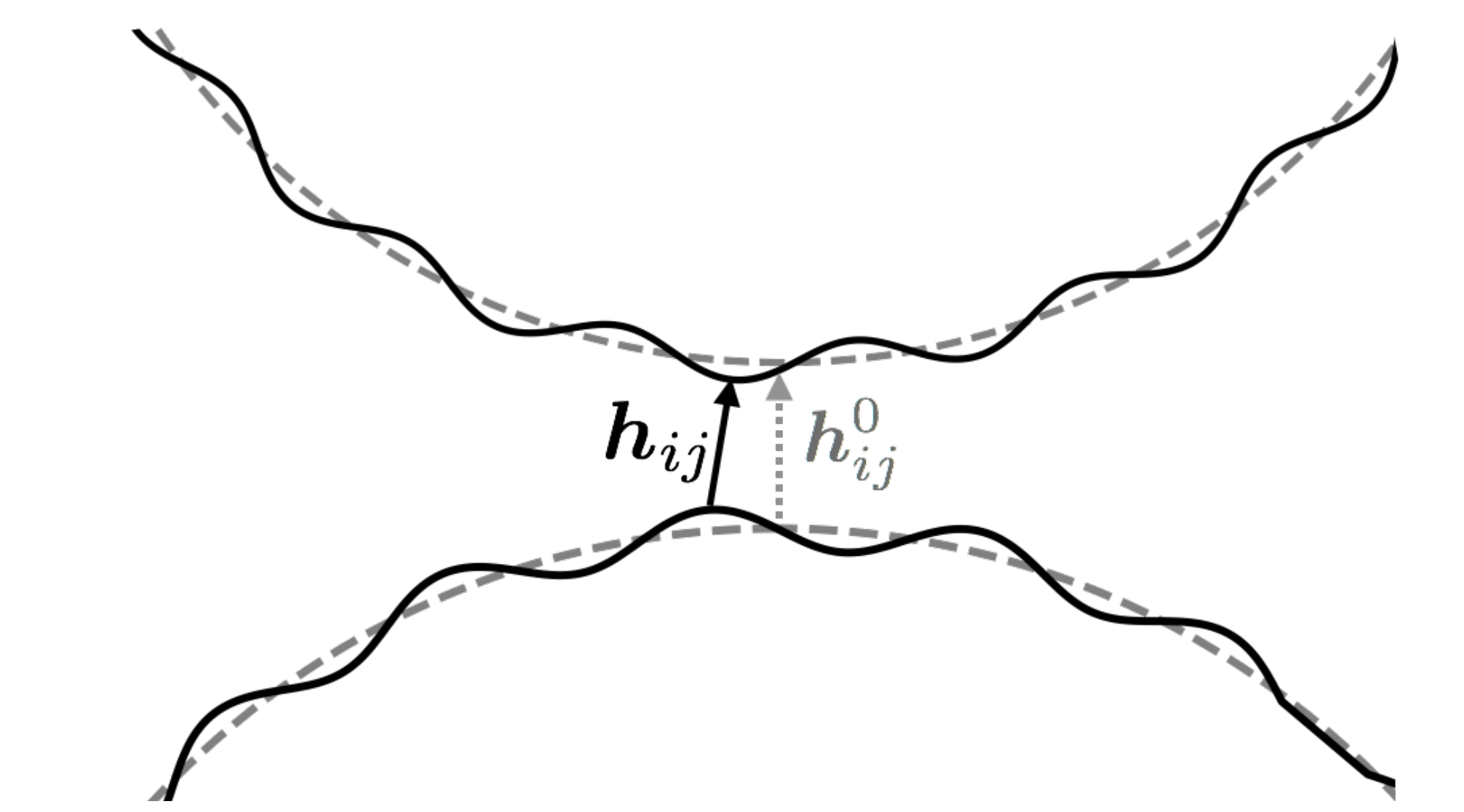} \caption{ Schematic
 picture of the surfaces of two particles. The solid lines represent the
 surfaces of particles, while the dashed lines represent the surfaces of
 the references disks. The solid and dashed arrows represent the minimal
 paths connecting the surfaces of particles and reference disks,
 respectively. } \label{124049_5Jan20}
\end{figure}
We consider two dimensional particles interacting
with the repulsive harmonic potential~\cite{ohern2003}:
\begin{align}
 V_N = \sum_{i<j}^{1,N} v(h_{ij}(\bx_i,\bx_j,u_i,u_j)),\
 v(h) = \frac{h^2}{2}\Theta(-h),
\end{align}
where $\Theta(x)$ denotes the Heaviside step function,
${\bx_i=\{x_i,y_i\}}$ and $u_i$ denote the position and angle of the
$i$-th particle, respectively, and $h_{ij}$ denotes the overlap function,
which represents the minimal distance between particles $i$ and $j$. When
particles $i$ and $j$ are overlapped $h_{ij}\leq 0$, and otherwise
$h_{ij}>0$. The calculation of $h_{ij}$ is a non-trivial task for
general shapes of non-spherical particles. To simplify the treatment, we
assume that the shape of the particles is close to a disk.  By means of a
perturbation expansion around the reference disks, we obtain~\cite{brito2018,ikeda2020}
\begin{align}
 h_{ij} &= \abs{\bh_{ij}} = \abs{\bh_{ij}^{0}}
 + \delta\bh_{ij}\cdot\frac{\bx_i-\bx_j}{\abs{\bx_i-\bx_j}}+O(\delta\bh_{ij}^2)\new
 &\approx \abs{\bx_i-\bx_j} 
 -R_i-R_j  + F(\bx_i,\bx_j,u_i,u_j),
\end{align}
where $\bh_{ij}$ and $\bh_{ij}^0$ respectively denote the vectors
connecting the minimal paths between the surfaces of two particles and
reference disks (Fig.~\ref{124049_5Jan20}), $\delta\bh_{ij}=\bh_{ij} -
\bh_{ij}^0$ denotes the deviation of the minimal path from the disks,
$R_i$ denotes the radius of the particles $i$, and we have introduced
the auxiliary function $F\equiv
\delta\bh_{ij}\cdot(\bx_i-\bx_j)/\abs{\bx_i-\bx_j}$.  To express the
surface roughness, we require that $F$ is invariant under the following
transformations: (i) the rotation without slip ${u_i\to
u_i+\delta/R_i}$ and ${u_j\to u_j-\delta/R_j}$, where $\delta$ denotes
an arbitrary constant with a dimension of length, and (ii) the
global rotation. A functional form satisfying the above conditions
is~\footnote{$F$ is not a periodic function of $u_i$ and $u_j$. But,
this is not a big problem because such symmetries do not affect the
local stability arguments discussed here.}
\begin{align}
 &F(\bx_i,\bx_j,u_i,u_j) = (R_i+R_j)f\left(\omega_{ij}\right),\new
 &\omega_{ij} = \frac{R_iu_i + R_j u_j}{R_i+R_j}-\theta_{ij},
 \label{145744_2Jan20} 
\end{align}
where $\theta_{ij}$ denotes the angle between the relative vector
$\bx_i-\bx_j$ and positive $x$-axis, namely,
${\theta_{ij}=\atan2(y_j-y_i,x_j-x_i)}$.
\begin{figure} 
 \centering
 \includegraphics[width=8.5cm]{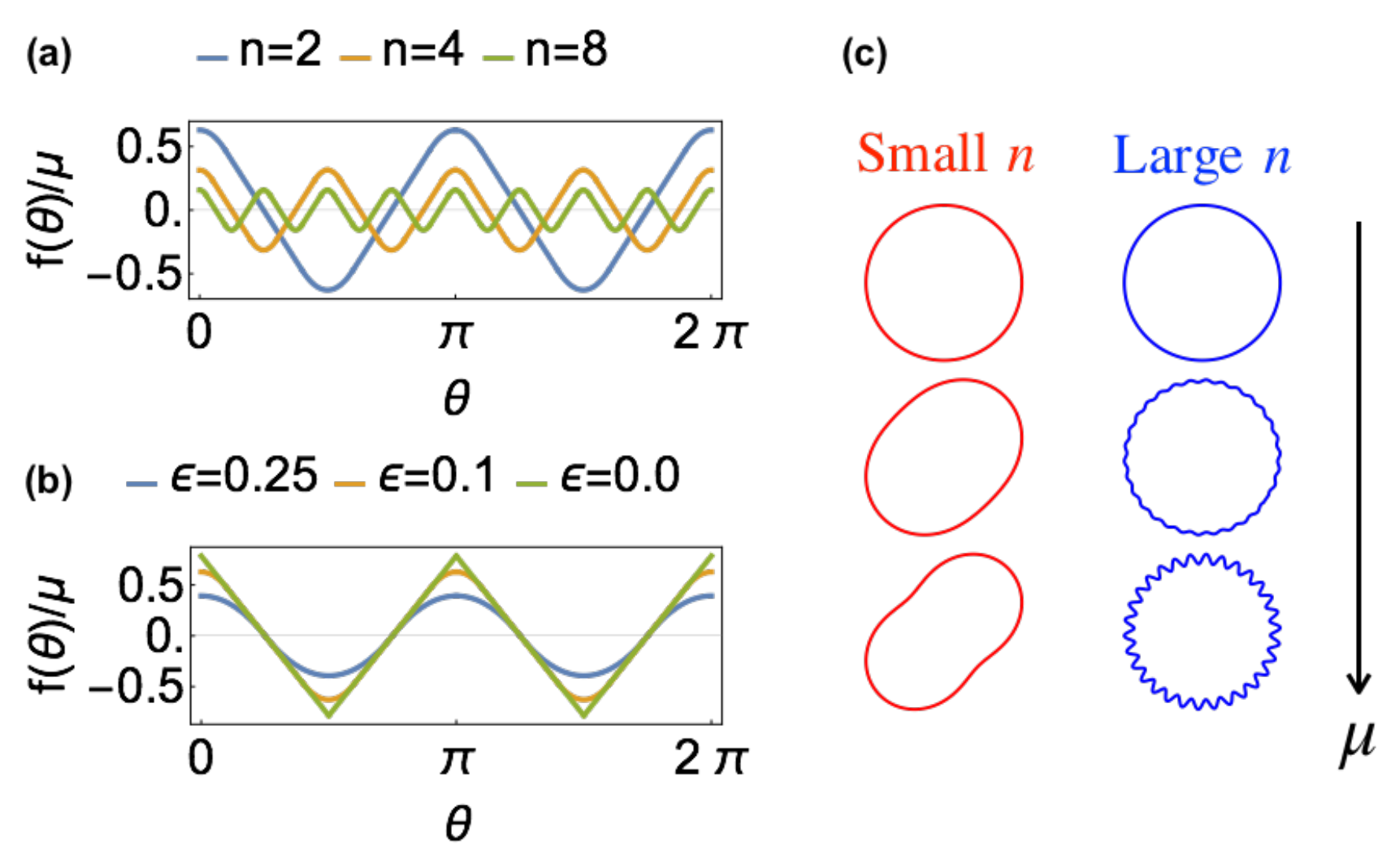}

 \caption{ (a) $f(\theta)/\mu$ for $\varepsilon=0.1$ and several $n$. The
number of minima increases with $n$. (b) $f(\theta)$ for $n=2$ and
several $\varepsilon$. The triangle wave is recovered when $\varepsilon
= 0$. (c) Schematic pictures of a particle shape. The shape
deviates from the reference disk with $\mu$, and the surface becomes
rougher for larger $n$.}  \label{121428_16Jan20}
\end{figure}
Although $f(\theta)$ can be any periodic function of period $\pi$,
to make the connection with the Coulomb friction law, we
consider the following specific form:
\begin{align}
 f(\theta) = \mu\frac{2\pi }{n}\tri\left(\frac{n}{2\pi}\theta\right),\label{162355_2Jan20}
\end{align}
where $n$ denotes an even number, and we have introduced a smoothed
triangle wave function:
\begin{align}
 \tri(x) =
 \begin{cases}
  -\frac{x^2}{2\varepsilon} +\frac{1}{4}-\frac{\varepsilon}{2}, & x\in[0,\varepsilon),\\
  -x +\frac{1}{4}, & x\in[\varepsilon,1/2-\varepsilon),\\
  \frac{(x-1/2)^2}{2\varepsilon} -\frac{1}{4}+\frac{\varepsilon}{2}, & x\in[1/2-\varepsilon,1/2+\varepsilon),\\
  x -\frac{3}{4}, & x\in[1/2+\varepsilon,1-\varepsilon),\\
  -\frac{(x-1)^2}{2\varepsilon} +\frac{1}{4}-\frac{\varepsilon}{2}, & x\in[1-\varepsilon,1),
 \end{cases}
\end{align}
and $\tri(x\pm 1)=\tri(x)$. We show the typical behavior of $f(\theta)$
in Figs.~\ref{121428_16Jan20}(a,b).  $f(\theta)$ depends on three
parameters: $n$, $\mu$, and $\varepsilon$.  Upon increasing $n$, the
number of minima of $f(\theta)$ increases. Although solely $f(\theta)$
is not enough to determine the precise shape of particles, it is clear
that the number of minima on the surface of a particle also increases
with $n$, as schematically shown in Fig~\ref{121428_16Jan20}(c),
suggesting that $n$ controls the roughness.  $\mu/n$ represents the
deviation from the reference disks, and our perturbative approach is
justified only for ${\mu/n\ll 1}$. To make the physical meaning of $\mu$
more clear, we calculate the ratio between the normal and tangential
forces between two particles in contact:
\begin{align}
 \abs{\frac{f_t}{f_n}} = \frac{R_i+R_j}{\abs{\bx_i-\bx_j}}\abs{f'\left(\omega_{ij}\right)},
\end{align}
where $f_n=\partial_{x_n}v(h_{ij})$ and $f_t=\partial_{x_t}v(h_{ij})$.
$\partial_{x_n}$ and $\partial_{x_t}$ respectively denote the
derivatives along the parallel and orthogonal directions to
$\bm{x}_i-\bm{x}_j$. For $n\gg 1$ and at $\varphi_J$,
$\abs{\bx_i-\bx_j}\approx R_i + R_j$, and we get $\abs{f_t/f_t}\approx
f'(\omega_{ij}) \leq \mu$, implying that $\mu$ represents the effective
friction coefficient. One may thus expect that the behavior of
frictional particles can be recovered in the limit of a rough
surface, $n\to \infty$ with fixed $\mu$. However, taking this limit is
not enough because for $\varepsilon>0$, $\abs{f_t/f_n}=f'(\omega_{ij})$
varies depending on $\omega_{ij}$ even when slip sets in, while
$\abs{f_t/f_n}=\mu$ for the Coulomb friction law~\cite{cundall1979}.
The Coulomb friction law corresponds to the double limit $n\to\infty$
and $\varepsilon\to 0$, where $\abs{f_t/f_n}<\mu$ if $\omega_{ij}$ is
trapped in a minimum of $f(\omega_{ij})$, and $\abs{f_t/f_n}=\mu$ 
if slip sets in and $\omega_{ij} \in
[\varepsilon,1/2-\varepsilon)\cup [1/2+\varepsilon,1-\varepsilon)$.

\paragraph*{Numerics. --}

We perform numerical simulations for $N=128$ particles consisting of the
same number of large and small particles under periodic boundary
conditions. The radii of small and large particles are $R_S=0.5$ and
$R_L=0.7$, respectively. We find $\varphi_J$ by combining slow isotropic
compression and decompression as follows~\cite{ohern2003}. We first
generate a random initial configuration at a small packing fraction
$\varphi=0.1$. Then, we slowly compress the system. For each compression
step, we increase the packing fraction with a small increment $\delta
\varphi=10^{-4}$, and successively minimize the energy with the FIRE
algorithm~\cite{fire2006} until the squared force acting on each
particle becomes smaller than $10^{-25}$. After arriving at a jammed
configuration with $V_N/N > 10^{-16}$, we change the sign and amplitude
of the increment as $\delta\varphi\to -\delta\varphi/2$. Then, we
decompress the system until we obtain an unjammed configuration with
$V_N/N < 10^{-16}$. We repeat this process by changing the sign and
amplitude of the increment as $\delta\varphi\to -\delta\varphi/2$ every
time the system crosses $\varphi_J$. We terminate the
simulation when $V_N/N\in (10^{-16},2\times 10^{-16})$. Then, we remove
the rattlers that have zero or one contact, and calculate the physical
quantities.  To improve the statistics, we average over $10$ independent
samples.

\paragraph*{Results. --}

\begin{figure} 
 \centering \includegraphics[width=9cm]{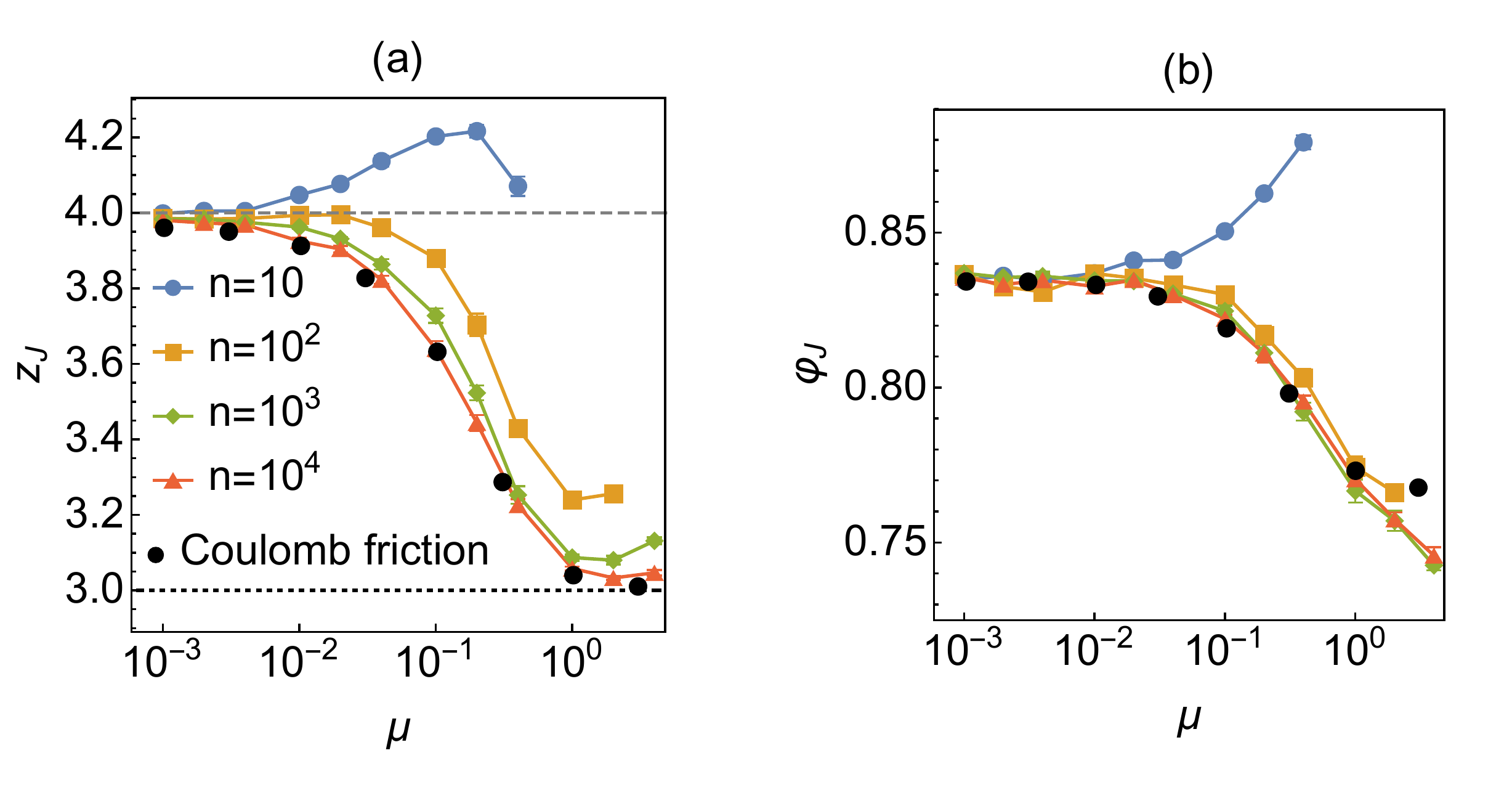} \caption{$\mu$ dependence for fixed $\varepsilon=0.1$ of (a) the contact
 number per particle at the jamming transition point $z_J$ and (b) the jamming transition point $\varphi_J$. The
 data for the Coulomb friction model were taken from
 Ref.~\cite{xiong2019}.}  \label{132556_2Jan20}
\end{figure}
First, we discuss the behavior for $\varepsilon=0.1$. In
Fig.~\ref{132556_2Jan20}(a), we show the contact number per particle at
the jamming transition point $z_J$. For small $n$, $z_J$ increases upon
increasing $\mu$, see the data for $n=10$. Since $\mu/n$ represents the
deviation from disks, this behavior is qualitatively similar to that
observed in convex-shaped
particles~\cite{donev2004,donev2007,williams2003,blouwolff2006,werf2018,brito2018,ikeda2019mean,ikeda2020}.
Contrarily, for large $n$, $z_J$ decreases with $\mu$~\footnote{ For
large $n$, $z_J$ has a minimum at an intermediate value of $\mu$. This
counter-intuitive behavior could be a signature of the breakdown of our
perturbative approach. The position of the minimum shifts to higher
$\mu$ upon increasing of $n$, and the result of the Coulomb friction law
is correctly reproduced in the limit of $n\to\infty$.}. For the largest
value of $n$, $n=10^4$, $z_J$ quantitatively agrees with previous
results generated by isotropic compression of frictional particles for
the same system size $N=128$~\cite{xiong2019}. In
Fig.~\ref{132556_2Jan20}(b), we show the jamming transition point
$\varphi_J$. As for $z_J$, $\varphi_J$ increases with $\mu$ for small
$n$, and decreases for large $n$. For $n=10^4$ and $\mu\lesssim 1$, the
behavior of $\varphi_J$ is similar to that of the Coulomb friction
model, while for $\mu\gtrsim 1$, there is a small but visible
deviation. We guess that this discrepancy for large $\mu$ is due to the
difference in the algorithms used for the minimization: for our model,
the energy was minimized by the FIRE algorithm, while for frictional
particles, the kinetic energy was minimized by molecular dynamics
simulation with a damping proportional to the
force~\cite{xiong2019}. Further studies are necessary to clarify this
point.
\begin{figure} 
 \centering \includegraphics[width=9cm]{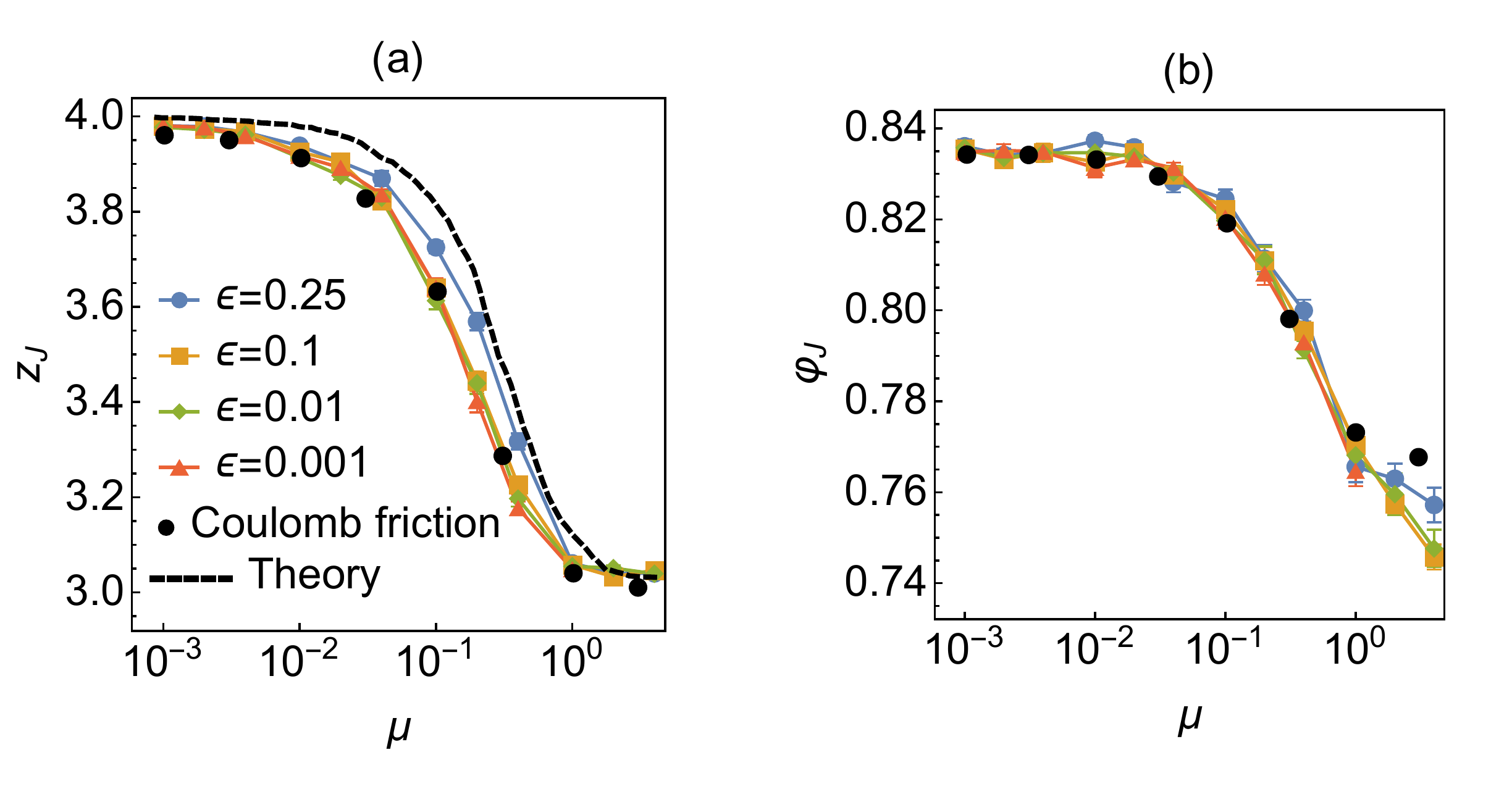} \caption{ $\mu$
 dependence for fixed $n=10^4$ of (a) the contact number per particle at
 the jamming transition point $z_J$ and (b) the jamming transition point
 $\varphi_J$.  Markers denote the numerical results, while the dashed
 line denotes the theoretical prediction, see main text. }
 \label{181602_2Jan20}
 \end{figure}
To see the $\varepsilon$ dependence, in Fig.~\ref{181602_2Jan20} we
show $z_J$ and $\varphi_J$ for $n=10^4$ and several $\varepsilon$. $z_J$
and $\varphi_J$ do not exhibit a significant $\varepsilon$ dependence
and agree well with the results for the Coulomb friction law.

\begin{figure*}
 \centering \includegraphics[width=18cm]{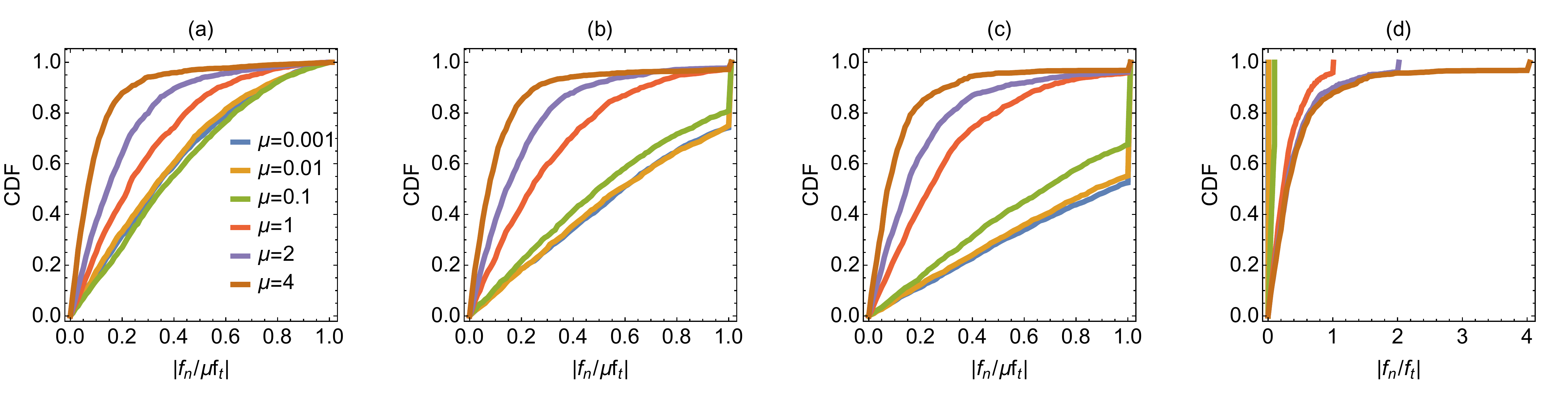} \caption{ (a)--(c)
 Cumulative distributions of $\abs{f_n/\mu f_t}$ for $n=10^4$ and
 $\varepsilon=0.25$, $0.1$, and $0.01$, respectively.  (d) Cumulative
 distribution of $\abs{f_n/f_t}$ for $n=10^4$ and $\varepsilon=0.01$.
}
						 \label{100425_2Jan20}
\end{figure*}
In Fig.~\ref{100425_2Jan20}(a)--(c), we show the cumulative
distribution function (CDF) of $\abs{f_n/\mu f_t}$ for $n=10^4$ and
several $\varepsilon$. For $\varepsilon=0.25$, the CDF smoothly
increases with $\abs{f_n/\mu f_t}$. Contrarily, for
$\varepsilon<0.25$, the CDF has a singular peak at $\abs{f_n/\mu
f_t}=1$. The peak grows upon decreasing $\mu$ and $\varepsilon$. In
Fig.~\ref{100425_2Jan20}(d), we show the CDF of $\abs{f_n/f_t}$ for $n=10^4$
and $\varepsilon=0.01$. The CDF converges to a constant distribution for
large $\mu$.

\begin{figure}
 \centering \includegraphics[width=7cm]{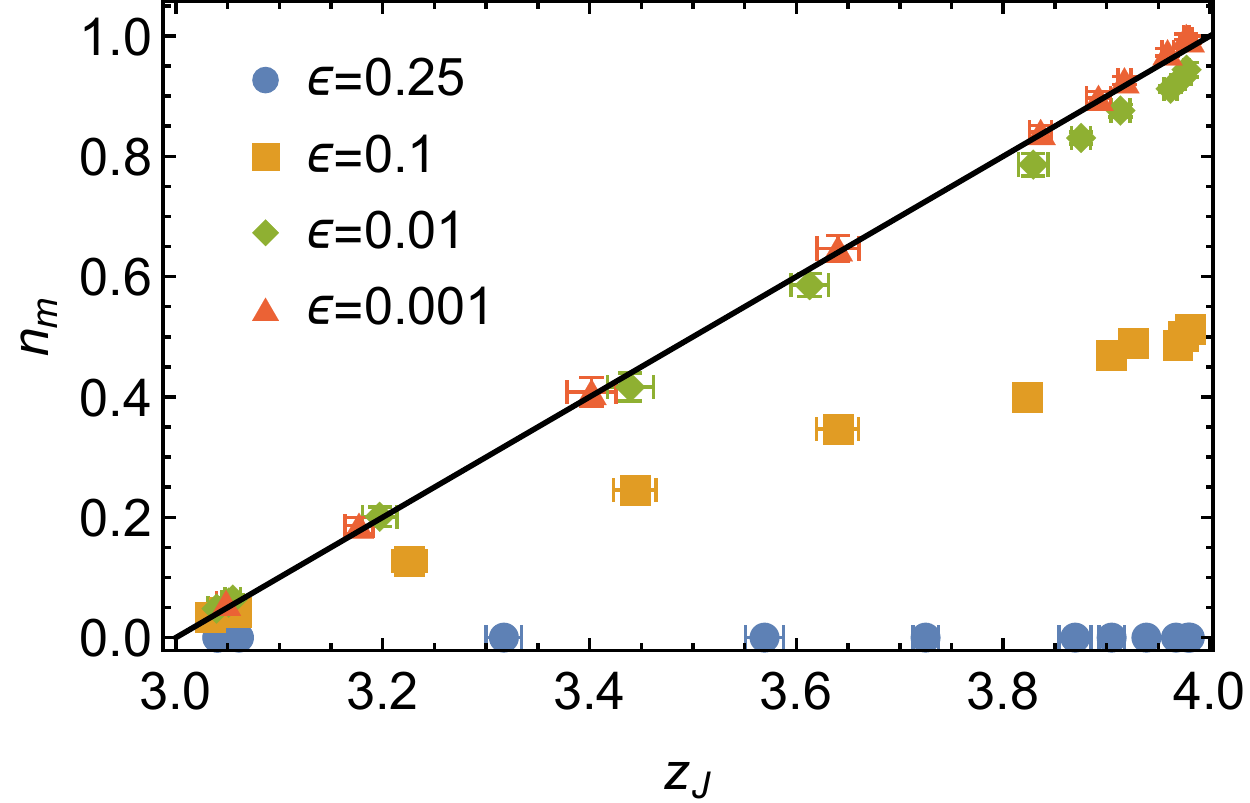} \caption{
 Relation between $n_m$ and $z_J$ for $n=10^4$. Markers denote numerical
 results.  The solid line denotes the theoretical prediction $n_m=z-3$
corresponding to generalized isostaticity.}
 \label{121440_2Jan20}
\end{figure}
The strong peak of the CDF at $\abs{f_t/\mu
f_n}=1$ indicates that there are a finite fraction of contacts
satisfying the Coulomb threshold $\abs{f_t} = \mu f_n$. Those contacts
are referred to as the fully mobilized contacts~\cite{van2009}. As the
fully mobilized contacts do not constrain the tangential motion, the
total number of constraints imposed by the contacts is $Nz-Nn_m$, where
$Nn_m$ denotes the number of fully mobilized contacts. This should be
equated to the number of degrees of freedom $3N$ when the system is
isostatic. Therefore, for an isostatic system, $n_m$
is~\cite{henkes2010}
\begin{align}
 n_m = z-3.\label{121730_2Jan20}
\end{align}
In Fig.~\ref{121440_2Jan20}, we test this conjecture for $n=10^4$. The
plot clearly shows that the numerical data converge to the theoretical
prediction, Eq.~(\ref{121730_2Jan20}), in the Coulomb
friction limit $\varepsilon\to 0$.

\paragraph*{Theory. --}
Here we show that the generalized isostaticity in the $\varepsilon \to
0$ limit can be explained by a simple counting argument, which slightly
generalizes the corresponding one for nonspherical particles in two
dimensions~\cite{van2009}. At $\varphi_J$, $h_{ij}=0$
for all contacts. This can be satisfied when the number of degrees of
freedom $3N$ is larger than the number of contacts $Nz_J/2$:
\begin{align}
 3N \geq \frac{Nz_J}{2}\to z_J\leq 6.\label{101433_6Jan20}
\end{align}
Besides, a stable system should
satisfy the $3N$ force balance equations:
\begin{align}
 \pdiff{V_N}{x_i}=0,\ \pdiff{V_N}{y_i}=0,\ \pdiff{V_N}{u_i}=0.\label{101258_6Jan20}
\end{align}
Those are linear combinations of the normal and tangential forces, $f_n$
and $f_t$, between particles in contact. Considering that there are
$Nn_m$ fully mobilized contacts, the degrees of freedom of $f_n$ and
$f_t$ is $Nz_J-Nn_m$. Therefore, Eqs.~(\ref{101258_6Jan20}) have
non-trivial solutions only if~\cite{shundyak2007,henkes2010}
\begin{align}
 3N\leq Nz_J -Nn_m \to z_J\geq 3+n_m.\label{101437_6Jan20}
\end{align}
This inequality generally holds for any $\varepsilon$, see
Fig.~\ref{121440_2Jan20}. From Eqs.~(\ref{101433_6Jan20}) and
(\ref{101437_6Jan20}), at $\varphi_J$, we have
\begin{align}
 3+n_m\leq z_J \leq 6,
\end{align}
implying that the generalized isostaticity does not hold in general.

We can improve the counting argument in the $\varepsilon\to 0$
limit, where $\omega_{ij}$ of non-mobilized contacts should be
located precisely at a minimum of $f(\omega_{ij})$ because the
corresponding stiffness diverges as $f''(\omega_{ij})\sim
\varepsilon^{-1}$. This provides $Nz_J/2-Nn_m$ additional
constraints. Thus, Eq.~(\ref{101433_6Jan20}) should be modified as
\begin{align}
 3N \geq Nz_J-Nn_m \to z_J\leq 3+n_m.\label{105837_6Jan20}
\end{align}
Eqs.~(\ref{101437_6Jan20}) and (\ref{105837_6Jan20}) prove the
generalized isostaticity Eq.~(\ref{121730_2Jan20}).  It is not
straightforward to generalize the above argument to higher
dimensions. We left it as future work.

A theoretical challenge is to predict the $\mu$ dependence of $z_J$.
In Fig.~\ref{181602_2Jan20}(a), we show that $z_J$ does not show a strong
$\varepsilon$ dependence. This allows us to focus on small
$\varepsilon$, e.g. $\varepsilon=0.01$, where the generalized isostaticity,
Eq.~(\ref{121730_2Jan20}), may simplify the treatment, as it directly
connects $z_J$ to $n_m$.  We tentatively approximate $n_m$ by
neglecting the $\mu$ dependence of the distribution of $\abs{f_t/f_n}$ and
calculate $n_m$ as
\begin{align}
 n_m \approx \int_\mu^\infty dx P_{\infty}(x)
 = 1-{\rm CDF}(\abs{f_t/f_n}=\mu),\label{143726_10Jan20}
\end{align}
where $P_\infty(x)$ denotes the distribution of $\abs{f_t/f_n}$ in the
limit $\mu\to\infty$. As shown in Fig.~\ref{100425_2Jan20}(d), the CDF
of $\abs{f_t/f_n}$ for $\varepsilon=0.01$ is converged to a constant
distribution for $\mu \gtrsim 2$. So, we use the CDF for $\mu=4$. In
Fig.~\ref{181602_2Jan20}(a), we show our theoretical prediction
$z_J\approx n_m + 3 \approx 4-{\rm CDF}(\abs{f_t/f_n}=\mu)$ with the
black dashed line. The agreement is not perfect but still surprisingly
nice, considering the simplicity of the theory and fact that there are
only a few theories for the jamming of frictional
particles~\cite{song2008phase,shen2014}.

\paragraph*{Conclusions. --}
We constructed a model that takes into account the effect
of surface roughness by means of a perturbation expansion around ideal disks.  By
changing the surface roughness, the model can smoothly interpolate the
phenomenology of frictionless convex-shaped particles and frictional
disks.

We found that the fraction of fully mobilized contacts strongly
depends on $\varepsilon$, and, consequently, the generalized
isostaticity condition is satisfied only in the limit of the Coulomb friction law,
$\varepsilon\to 0$. However, our investigation is limited to a
specific class of functions $f(\omega_{ij})$ described by
Eq.~(\ref{162355_2Jan20}), and we also assumed that two particles have
at most one contact and neglected the effect of multiple contacts. It would be
desirable to investigate a broader class of $f(\omega_{ij})$,
allowing multiple contacts, to clarify under which conditions the systems satisfies
generalized isostaticity.

\begin{acknowledgments}
\paragraph*{Acknowledgements. --}
We warmly thank J.-P.~Bouchaud for discussions related to this work.
This project has received funding from the European Research Council
(ERC) under the European Union's Horizon 2020 research and innovation
program (grant agreement n.~723955-GlassUniversality).
\end{acknowledgments}


\bibliography{apssamp}

\begin{thebibliography}{44}%
\makeatletter
\providecommand \@ifxundefined [1]{%
 \@ifx{#1\undefined}
}%
\providecommand \@ifnum [1]{%
 \ifnum #1\expandafter \@firstoftwo
 \else \expandafter \@secondoftwo
 \fi
}%
\providecommand \@ifx [1]{%
 \ifx #1\expandafter \@firstoftwo
 \else \expandafter \@secondoftwo
 \fi
}%
\providecommand \natexlab [1]{#1}%
\providecommand \enquote  [1]{``#1''}%
\providecommand \bibnamefont  [1]{#1}%
\providecommand \bibfnamefont [1]{#1}%
\providecommand \citenamefont [1]{#1}%
\providecommand \href@noop [0]{\@secondoftwo}%
\providecommand \href [0]{\begingroup \@sanitize@url \@href}%
\providecommand \@href[1]{\@@startlink{#1}\@@href}%
\providecommand \@@href[1]{\endgroup#1\@@endlink}%
\providecommand \@sanitize@url [0]{\catcode `\\12\catcode `\$12\catcode
  `\&12\catcode `\#12\catcode `\^12\catcode `\_12\catcode `\%12\relax}%
\providecommand \@@startlink[1]{}%
\providecommand \@@endlink[0]{}%
\providecommand \url  [0]{\begingroup\@sanitize@url \@url }%
\providecommand \@url [1]{\endgroup\@href {#1}{\urlprefix }}%
\providecommand \urlprefix  [0]{URL }%
\providecommand \Eprint [0]{\href }%
\providecommand \doibase [0]{http://dx.doi.org/}%
\providecommand \selectlanguage [0]{\@gobble}%
\providecommand \bibinfo  [0]{\@secondoftwo}%
\providecommand \bibfield  [0]{\@secondoftwo}%
\providecommand \translation [1]{[#1]}%
\providecommand \BibitemOpen [0]{}%
\providecommand \bibitemStop [0]{}%
\providecommand \bibitemNoStop [0]{.\EOS\space}%
\providecommand \EOS [0]{\spacefactor3000\relax}%
\providecommand \BibitemShut  [1]{\csname bibitem#1\endcsname}%
\let\auto@bib@innerbib\@empty
\bibitem [{\citenamefont {Van~Hecke}(2009)}]{van2009}%
  \BibitemOpen
  \bibfield  {author} {\bibinfo {author} {\bibfnamefont {M.}~\bibnamefont
  {Van~Hecke}},\ }\href@noop {} {\bibfield  {journal} {\bibinfo  {journal}
  {Journal of Physics: Condensed Matter}\ }\textbf {\bibinfo {volume} {22}},\
  \bibinfo {pages} {033101} (\bibinfo {year} {2009})}\BibitemShut {NoStop}%
\bibitem [{\citenamefont {Durian}(1995)}]{durian1995}%
  \BibitemOpen
  \bibfield  {author} {\bibinfo {author} {\bibfnamefont {D.~J.}\ \bibnamefont
  {Durian}},\ }\href {\doibase 10.1103/PhysRevLett.75.4780} {\bibfield
  {journal} {\bibinfo  {journal} {Phys. Rev. Lett.}\ }\textbf {\bibinfo
  {volume} {75}},\ \bibinfo {pages} {4780} (\bibinfo {year}
  {1995})}\BibitemShut {NoStop}%
\bibitem [{\citenamefont {Zhang}\ \emph {et~al.}(2009)\citenamefont {Zhang},
  \citenamefont {Xu}, \citenamefont {Chen}, \citenamefont {Yunker},
  \citenamefont {Alsayed}, \citenamefont {Aptowicz}, \citenamefont {Habdas},
  \citenamefont {Liu}, \citenamefont {Nagel},\ and\ \citenamefont
  {Yodh}}]{zhang2009}%
  \BibitemOpen
  \bibfield  {author} {\bibinfo {author} {\bibfnamefont {Z.}~\bibnamefont
  {Zhang}}, \bibinfo {author} {\bibfnamefont {N.}~\bibnamefont {Xu}}, \bibinfo
  {author} {\bibfnamefont {D.~T.}\ \bibnamefont {Chen}}, \bibinfo {author}
  {\bibfnamefont {P.}~\bibnamefont {Yunker}}, \bibinfo {author} {\bibfnamefont
  {A.~M.}\ \bibnamefont {Alsayed}}, \bibinfo {author} {\bibfnamefont {K.~B.}\
  \bibnamefont {Aptowicz}}, \bibinfo {author} {\bibfnamefont {P.}~\bibnamefont
  {Habdas}}, \bibinfo {author} {\bibfnamefont {A.~J.}\ \bibnamefont {Liu}},
  \bibinfo {author} {\bibfnamefont {S.~R.}\ \bibnamefont {Nagel}}, \ and\
  \bibinfo {author} {\bibfnamefont {A.~G.}\ \bibnamefont {Yodh}},\ }\href@noop
  {} {\bibfield  {journal} {\bibinfo  {journal} {Nature}\ }\textbf {\bibinfo
  {volume} {459}},\ \bibinfo {pages} {230} (\bibinfo {year}
  {2009})}\BibitemShut {NoStop}%
\bibitem [{\citenamefont {Bernal}\ and\ \citenamefont
  {Mason}(1960)}]{bernal1960}%
  \BibitemOpen
  \bibfield  {author} {\bibinfo {author} {\bibfnamefont {J.}~\bibnamefont
  {Bernal}}\ and\ \bibinfo {author} {\bibfnamefont {J.}~\bibnamefont {Mason}},\
  }\href@noop {} {\bibfield  {journal} {\bibinfo  {journal} {Nature}\ }\textbf
  {\bibinfo {volume} {188}},\ \bibinfo {pages} {910} (\bibinfo {year}
  {1960})}\BibitemShut {NoStop}%
\bibitem [{\citenamefont {Donev}\ \emph {et~al.}(2004)\citenamefont {Donev},
  \citenamefont {Cisse}, \citenamefont {Sachs}, \citenamefont {Variano},
  \citenamefont {Stillinger}, \citenamefont {Connelly}, \citenamefont
  {Torquato},\ and\ \citenamefont {Chaikin}}]{donev2004}%
  \BibitemOpen
  \bibfield  {author} {\bibinfo {author} {\bibfnamefont {A.}~\bibnamefont
  {Donev}}, \bibinfo {author} {\bibfnamefont {I.}~\bibnamefont {Cisse}},
  \bibinfo {author} {\bibfnamefont {D.}~\bibnamefont {Sachs}}, \bibinfo
  {author} {\bibfnamefont {E.~A.}\ \bibnamefont {Variano}}, \bibinfo {author}
  {\bibfnamefont {F.~H.}\ \bibnamefont {Stillinger}}, \bibinfo {author}
  {\bibfnamefont {R.}~\bibnamefont {Connelly}}, \bibinfo {author}
  {\bibfnamefont {S.}~\bibnamefont {Torquato}}, \ and\ \bibinfo {author}
  {\bibfnamefont {P.~M.}\ \bibnamefont {Chaikin}},\ }\href@noop {} {\bibfield
  {journal} {\bibinfo  {journal} {Science}\ }\textbf {\bibinfo {volume}
  {303}},\ \bibinfo {pages} {990} (\bibinfo {year} {2004})}\BibitemShut
  {NoStop}%
\bibitem [{\citenamefont {Jaoshvili}\ \emph {et~al.}(2010)\citenamefont
  {Jaoshvili}, \citenamefont {Esakia}, \citenamefont {Porrati},\ and\
  \citenamefont {Chaikin}}]{jaoshvili2010}%
  \BibitemOpen
  \bibfield  {author} {\bibinfo {author} {\bibfnamefont {A.}~\bibnamefont
  {Jaoshvili}}, \bibinfo {author} {\bibfnamefont {A.}~\bibnamefont {Esakia}},
  \bibinfo {author} {\bibfnamefont {M.}~\bibnamefont {Porrati}}, \ and\
  \bibinfo {author} {\bibfnamefont {P.~M.}\ \bibnamefont {Chaikin}},\ }\href
  {\doibase 10.1103/PhysRevLett.104.185501} {\bibfield  {journal} {\bibinfo
  {journal} {Phys. Rev. Lett.}\ }\textbf {\bibinfo {volume} {104}},\ \bibinfo
  {pages} {185501} (\bibinfo {year} {2010})}\BibitemShut {NoStop}%
\bibitem [{\citenamefont {Bi}\ \emph {et~al.}(2015)\citenamefont {Bi},
  \citenamefont {Lopez}, \citenamefont {Schwarz},\ and\ \citenamefont
  {Manning}}]{bi2015}%
  \BibitemOpen
  \bibfield  {author} {\bibinfo {author} {\bibfnamefont {D.}~\bibnamefont
  {Bi}}, \bibinfo {author} {\bibfnamefont {J.}~\bibnamefont {Lopez}}, \bibinfo
  {author} {\bibfnamefont {J.}~\bibnamefont {Schwarz}}, \ and\ \bibinfo
  {author} {\bibfnamefont {M.~L.}\ \bibnamefont {Manning}},\ }\href@noop {}
  {\bibfield  {journal} {\bibinfo  {journal} {Nature Physics}\ }\textbf
  {\bibinfo {volume} {11}},\ \bibinfo {pages} {1074} (\bibinfo {year}
  {2015})}\BibitemShut {NoStop}%
\bibitem [{\citenamefont {Franz}\ and\ \citenamefont
  {Parisi}(2016)}]{franz2016}%
  \BibitemOpen
  \bibfield  {author} {\bibinfo {author} {\bibfnamefont {S.}~\bibnamefont
  {Franz}}\ and\ \bibinfo {author} {\bibfnamefont {G.}~\bibnamefont {Parisi}},\
  }\href@noop {} {\bibfield  {journal} {\bibinfo  {journal} {Journal of Physics
  A: Mathematical and Theoretical}\ }\textbf {\bibinfo {volume} {49}},\
  \bibinfo {pages} {145001} (\bibinfo {year} {2016})}\BibitemShut {NoStop}%
\bibitem [{\citenamefont {Franz}\ \emph
  {et~al.}(2019{\natexlab{a}})\citenamefont {Franz}, \citenamefont {Hwang},\
  and\ \citenamefont {Urbani}}]{franz2019multilayer}%
  \BibitemOpen
  \bibfield  {author} {\bibinfo {author} {\bibfnamefont {S.}~\bibnamefont
  {Franz}}, \bibinfo {author} {\bibfnamefont {S.}~\bibnamefont {Hwang}}, \ and\
  \bibinfo {author} {\bibfnamefont {P.}~\bibnamefont {Urbani}},\ }\href
  {\doibase 10.1103/PhysRevLett.123.160602} {\bibfield  {journal} {\bibinfo
  {journal} {Phys. Rev. Lett.}\ }\textbf {\bibinfo {volume} {123}},\ \bibinfo
  {pages} {160602} (\bibinfo {year} {2019}{\natexlab{a}})}\BibitemShut
  {NoStop}%
\bibitem [{\citenamefont {O'Hern}\ \emph {et~al.}(2003)\citenamefont {O'Hern},
  \citenamefont {Silbert}, \citenamefont {Liu},\ and\ \citenamefont
  {Nagel}}]{ohern2003}%
  \BibitemOpen
  \bibfield  {author} {\bibinfo {author} {\bibfnamefont {C.~S.}\ \bibnamefont
  {O'Hern}}, \bibinfo {author} {\bibfnamefont {L.~E.}\ \bibnamefont {Silbert}},
  \bibinfo {author} {\bibfnamefont {A.~J.}\ \bibnamefont {Liu}}, \ and\
  \bibinfo {author} {\bibfnamefont {S.~R.}\ \bibnamefont {Nagel}},\ }\href
  {\doibase 10.1103/PhysRevE.68.011306} {\bibfield  {journal} {\bibinfo
  {journal} {Phys. Rev. E}\ }\textbf {\bibinfo {volume} {68}},\ \bibinfo
  {pages} {011306} (\bibinfo {year} {2003})}\BibitemShut {NoStop}%
\bibitem [{\citenamefont {Goodrich}\ \emph {et~al.}(2012)\citenamefont
  {Goodrich}, \citenamefont {Liu},\ and\ \citenamefont
  {Nagel}}]{goodrich2012finite}%
  \BibitemOpen
  \bibfield  {author} {\bibinfo {author} {\bibfnamefont {C.~P.}\ \bibnamefont
  {Goodrich}}, \bibinfo {author} {\bibfnamefont {A.~J.}\ \bibnamefont {Liu}}, \
  and\ \bibinfo {author} {\bibfnamefont {S.~R.}\ \bibnamefont {Nagel}},\
  }\href@noop {} {\bibfield  {journal} {\bibinfo  {journal} {Physical review
  letters}\ }\textbf {\bibinfo {volume} {109}},\ \bibinfo {pages} {095704}
  (\bibinfo {year} {2012})}\BibitemShut {NoStop}%
\bibitem [{\citenamefont {Franz}\ \emph {et~al.}(2017)\citenamefont {Franz},
  \citenamefont {Parisi}, \citenamefont {Sevelev}, \citenamefont {Urbani},\
  and\ \citenamefont {Zamponi}}]{franz2017universality}%
  \BibitemOpen
  \bibfield  {author} {\bibinfo {author} {\bibfnamefont {S.}~\bibnamefont
  {Franz}}, \bibinfo {author} {\bibfnamefont {G.}~\bibnamefont {Parisi}},
  \bibinfo {author} {\bibfnamefont {M.}~\bibnamefont {Sevelev}}, \bibinfo
  {author} {\bibfnamefont {P.}~\bibnamefont {Urbani}}, \ and\ \bibinfo {author}
  {\bibfnamefont {F.}~\bibnamefont {Zamponi}},\ }\href@noop {} {\bibfield
  {journal} {\bibinfo  {journal} {SciPost Phys}\ }\textbf {\bibinfo {volume}
  {2}},\ \bibinfo {pages} {019} (\bibinfo {year} {2017})}\BibitemShut {NoStop}%
\bibitem [{\citenamefont {Franz}\ \emph
  {et~al.}(2019{\natexlab{b}})\citenamefont {Franz}, \citenamefont {Sclocchi},\
  and\ \citenamefont {Urbani}}]{franz2019linear}%
  \BibitemOpen
  \bibfield  {author} {\bibinfo {author} {\bibfnamefont {S.}~\bibnamefont
  {Franz}}, \bibinfo {author} {\bibfnamefont {A.}~\bibnamefont {Sclocchi}}, \
  and\ \bibinfo {author} {\bibfnamefont {P.}~\bibnamefont {Urbani}},\ }\href
  {\doibase 10.1103/PhysRevLett.123.115702} {\bibfield  {journal} {\bibinfo
  {journal} {Phys. Rev. Lett.}\ }\textbf {\bibinfo {volume} {123}},\ \bibinfo
  {pages} {115702} (\bibinfo {year} {2019}{\natexlab{b}})}\BibitemShut
  {NoStop}%
\bibitem [{\citenamefont {Wyart}(2005)}]{wyart2005rigidity}%
  \BibitemOpen
  \bibfield  {author} {\bibinfo {author} {\bibfnamefont {M.}~\bibnamefont
  {Wyart}},\ }\href@noop {} {\bibfield  {journal} {\bibinfo  {journal} {arXiv
  preprint cond-mat/0512155}\ } (\bibinfo {year} {2005})}\BibitemShut {NoStop}%
\bibitem [{\citenamefont {Wyart}\ \emph {et~al.}(2005)\citenamefont {Wyart},
  \citenamefont {Silbert}, \citenamefont {Nagel},\ and\ \citenamefont
  {Witten}}]{wyart2005compress}%
  \BibitemOpen
  \bibfield  {author} {\bibinfo {author} {\bibfnamefont {M.}~\bibnamefont
  {Wyart}}, \bibinfo {author} {\bibfnamefont {L.~E.}\ \bibnamefont {Silbert}},
  \bibinfo {author} {\bibfnamefont {S.~R.}\ \bibnamefont {Nagel}}, \ and\
  \bibinfo {author} {\bibfnamefont {T.~A.}\ \bibnamefont {Witten}},\ }\href
  {\doibase 10.1103/PhysRevE.72.051306} {\bibfield  {journal} {\bibinfo
  {journal} {Phys. Rev. E}\ }\textbf {\bibinfo {volume} {72}},\ \bibinfo
  {pages} {051306} (\bibinfo {year} {2005})}\BibitemShut {NoStop}%
\bibitem [{\citenamefont {DeGiuli}\ \emph {et~al.}(2014)\citenamefont
  {DeGiuli}, \citenamefont {Lerner}, \citenamefont {Brito},\ and\ \citenamefont
  {Wyart}}]{degiuli2014force}%
  \BibitemOpen
  \bibfield  {author} {\bibinfo {author} {\bibfnamefont {E.}~\bibnamefont
  {DeGiuli}}, \bibinfo {author} {\bibfnamefont {E.}~\bibnamefont {Lerner}},
  \bibinfo {author} {\bibfnamefont {C.}~\bibnamefont {Brito}}, \ and\ \bibinfo
  {author} {\bibfnamefont {M.}~\bibnamefont {Wyart}},\ }\href@noop {}
  {\bibfield  {journal} {\bibinfo  {journal} {Proceedings of the National
  Academy of Sciences}\ }\textbf {\bibinfo {volume} {111}},\ \bibinfo {pages}
  {17054} (\bibinfo {year} {2014})}\BibitemShut {NoStop}%
\bibitem [{\citenamefont {Charbonneau}\ \emph {et~al.}(2014)\citenamefont
  {Charbonneau}, \citenamefont {Kurchan}, \citenamefont {Parisi}, \citenamefont
  {Urbani},\ and\ \citenamefont {Zamponi}}]{charbonneau2014fractal}%
  \BibitemOpen
  \bibfield  {author} {\bibinfo {author} {\bibfnamefont {P.}~\bibnamefont
  {Charbonneau}}, \bibinfo {author} {\bibfnamefont {J.}~\bibnamefont
  {Kurchan}}, \bibinfo {author} {\bibfnamefont {G.}~\bibnamefont {Parisi}},
  \bibinfo {author} {\bibfnamefont {P.}~\bibnamefont {Urbani}}, \ and\ \bibinfo
  {author} {\bibfnamefont {F.}~\bibnamefont {Zamponi}},\ }\href@noop {}
  {\bibfield  {journal} {\bibinfo  {journal} {Nat. Commun.}\ }\textbf {\bibinfo
  {volume} {5}},\ \bibinfo {pages} {3725} (\bibinfo {year} {2014})}\BibitemShut
  {NoStop}%
\bibitem [{\citenamefont {Charbonneau}\ \emph {et~al.}(2015)\citenamefont
  {Charbonneau}, \citenamefont {Corwin}, \citenamefont {Parisi},\ and\
  \citenamefont {Zamponi}}]{charbonneau2015buckling}%
  \BibitemOpen
  \bibfield  {author} {\bibinfo {author} {\bibfnamefont {P.}~\bibnamefont
  {Charbonneau}}, \bibinfo {author} {\bibfnamefont {E.~I.}\ \bibnamefont
  {Corwin}}, \bibinfo {author} {\bibfnamefont {G.}~\bibnamefont {Parisi}}, \
  and\ \bibinfo {author} {\bibfnamefont {F.}~\bibnamefont {Zamponi}},\ }\href
  {\doibase 10.1103/PhysRevLett.114.125504} {\bibfield  {journal} {\bibinfo
  {journal} {Phys. Rev. Lett.}\ }\textbf {\bibinfo {volume} {114}},\ \bibinfo
  {pages} {125504} (\bibinfo {year} {2015})}\BibitemShut {NoStop}%
\bibitem [{\citenamefont {Cundall}\ and\ \citenamefont
  {Strack}(1979)}]{cundall1979}%
  \BibitemOpen
  \bibfield  {author} {\bibinfo {author} {\bibfnamefont {P.~A.}\ \bibnamefont
  {Cundall}}\ and\ \bibinfo {author} {\bibfnamefont {O.~D.}\ \bibnamefont
  {Strack}},\ }\href@noop {} {\bibfield  {journal} {\bibinfo  {journal}
  {geotechnique}\ }\textbf {\bibinfo {volume} {29}},\ \bibinfo {pages} {47}
  (\bibinfo {year} {1979})}\BibitemShut {NoStop}%
\bibitem [{\citenamefont {Edwards}\ and\ \citenamefont
  {Grinev}(1999)}]{ewdards1999}%
  \BibitemOpen
  \bibfield  {author} {\bibinfo {author} {\bibfnamefont {S.~F.}\ \bibnamefont
  {Edwards}}\ and\ \bibinfo {author} {\bibfnamefont {D.~V.}\ \bibnamefont
  {Grinev}},\ }\href {\doibase 10.1103/PhysRevLett.82.5397} {\bibfield
  {journal} {\bibinfo  {journal} {Phys. Rev. Lett.}\ }\textbf {\bibinfo
  {volume} {82}},\ \bibinfo {pages} {5397} (\bibinfo {year}
  {1999})}\BibitemShut {NoStop}%
\bibitem [{\citenamefont {Unger}\ \emph {et~al.}(2005)\citenamefont {Unger},
  \citenamefont {Kert\'esz},\ and\ \citenamefont {Wolf}}]{unger2005}%
  \BibitemOpen
  \bibfield  {author} {\bibinfo {author} {\bibfnamefont {T.}~\bibnamefont
  {Unger}}, \bibinfo {author} {\bibfnamefont {J.}~\bibnamefont {Kert\'esz}}, \
  and\ \bibinfo {author} {\bibfnamefont {D.~E.}\ \bibnamefont {Wolf}},\ }\href
  {\doibase 10.1103/PhysRevLett.94.178001} {\bibfield  {journal} {\bibinfo
  {journal} {Phys. Rev. Lett.}\ }\textbf {\bibinfo {volume} {94}},\ \bibinfo
  {pages} {178001} (\bibinfo {year} {2005})}\BibitemShut {NoStop}%
\bibitem [{\citenamefont {Silbert}(2010)}]{silbert2010jamming}%
  \BibitemOpen
  \bibfield  {author} {\bibinfo {author} {\bibfnamefont {L.~E.}\ \bibnamefont
  {Silbert}},\ }\href@noop {} {\bibfield  {journal} {\bibinfo  {journal} {Soft
  Matter}\ }\textbf {\bibinfo {volume} {6}},\ \bibinfo {pages} {2918} (\bibinfo
  {year} {2010})}\BibitemShut {NoStop}%
\bibitem [{\citenamefont {Bouchaud}(2002)}]{bouchaud2002granular}%
  \BibitemOpen
  \bibfield  {author} {\bibinfo {author} {\bibfnamefont {J.-P.}\ \bibnamefont
  {Bouchaud}},\ }\href@noop {} {\bibfield  {journal} {\bibinfo  {journal}
  {arXiv preprint cond-mat/0211196}\ } (\bibinfo {year} {2002})}\BibitemShut
  {NoStop}%
\bibitem [{\citenamefont {Shundyak}\ \emph {et~al.}(2007)\citenamefont
  {Shundyak}, \citenamefont {van Hecke},\ and\ \citenamefont {van
  Saarloos}}]{shundyak2007}%
  \BibitemOpen
  \bibfield  {author} {\bibinfo {author} {\bibfnamefont {K.}~\bibnamefont
  {Shundyak}}, \bibinfo {author} {\bibfnamefont {M.}~\bibnamefont {van Hecke}},
  \ and\ \bibinfo {author} {\bibfnamefont {W.}~\bibnamefont {van Saarloos}},\
  }\href {\doibase 10.1103/PhysRevE.75.010301} {\bibfield  {journal} {\bibinfo
  {journal} {Phys. Rev. E}\ }\textbf {\bibinfo {volume} {75}},\ \bibinfo
  {pages} {010301} (\bibinfo {year} {2007})}\BibitemShut {NoStop}%
\bibitem [{\citenamefont {Henkes}\ \emph {et~al.}(2010)\citenamefont {Henkes},
  \citenamefont {van Hecke},\ and\ \citenamefont {van Saarloos}}]{henkes2010}%
  \BibitemOpen
  \bibfield  {author} {\bibinfo {author} {\bibfnamefont {S.}~\bibnamefont
  {Henkes}}, \bibinfo {author} {\bibfnamefont {M.}~\bibnamefont {van Hecke}}, \
  and\ \bibinfo {author} {\bibfnamefont {W.}~\bibnamefont {van Saarloos}},\
  }\href@noop {} {\bibfield  {journal} {\bibinfo  {journal} {EPL (Europhysics
  Letters)}\ }\textbf {\bibinfo {volume} {90}},\ \bibinfo {pages} {14003}
  (\bibinfo {year} {2010})}\BibitemShut {NoStop}%
\bibitem [{\citenamefont {Chattoraj}\ \emph {et~al.}(2019)\citenamefont
  {Chattoraj}, \citenamefont {Gendelman}, \citenamefont {Pica~Ciamarra},\ and\
  \citenamefont {Procaccia}}]{chattoraj2019}%
  \BibitemOpen
  \bibfield  {author} {\bibinfo {author} {\bibfnamefont {J.}~\bibnamefont
  {Chattoraj}}, \bibinfo {author} {\bibfnamefont {O.}~\bibnamefont
  {Gendelman}}, \bibinfo {author} {\bibfnamefont {M.}~\bibnamefont
  {Pica~Ciamarra}}, \ and\ \bibinfo {author} {\bibfnamefont {I.}~\bibnamefont
  {Procaccia}},\ }\href {\doibase 10.1103/PhysRevLett.123.098003} {\bibfield
  {journal} {\bibinfo  {journal} {Phys. Rev. Lett.}\ }\textbf {\bibinfo
  {volume} {123}},\ \bibinfo {pages} {098003} (\bibinfo {year}
  {2019})}\BibitemShut {NoStop}%
\bibitem [{\citenamefont {Persson}(2013)}]{persson2013sliding}%
  \BibitemOpen
  \bibfield  {author} {\bibinfo {author} {\bibfnamefont {B.~N.}\ \bibnamefont
  {Persson}},\ }\href@noop {} {\emph {\bibinfo {title} {Sliding friction:
  physical principles and applications}}}\ (\bibinfo  {publisher} {Springer
  Science \& Business Media},\ \bibinfo {year} {2013})\BibitemShut {NoStop}%
\bibitem [{\citenamefont {Athanassiadis}\ \emph {et~al.}(2014)\citenamefont
  {Athanassiadis}, \citenamefont {Miskin}, \citenamefont {Kaplan},
  \citenamefont {Rodenberg}, \citenamefont {Lee}, \citenamefont {Merritt},
  \citenamefont {Brown}, \citenamefont {Amend}, \citenamefont {Lipson},\ and\
  \citenamefont {Jaeger}}]{athanassiadis2014particle}%
  \BibitemOpen
  \bibfield  {author} {\bibinfo {author} {\bibfnamefont {A.~G.}\ \bibnamefont
  {Athanassiadis}}, \bibinfo {author} {\bibfnamefont {M.~Z.}\ \bibnamefont
  {Miskin}}, \bibinfo {author} {\bibfnamefont {P.}~\bibnamefont {Kaplan}},
  \bibinfo {author} {\bibfnamefont {N.}~\bibnamefont {Rodenberg}}, \bibinfo
  {author} {\bibfnamefont {S.~H.}\ \bibnamefont {Lee}}, \bibinfo {author}
  {\bibfnamefont {J.}~\bibnamefont {Merritt}}, \bibinfo {author} {\bibfnamefont
  {E.}~\bibnamefont {Brown}}, \bibinfo {author} {\bibfnamefont
  {J.}~\bibnamefont {Amend}}, \bibinfo {author} {\bibfnamefont
  {H.}~\bibnamefont {Lipson}}, \ and\ \bibinfo {author} {\bibfnamefont {H.~M.}\
  \bibnamefont {Jaeger}},\ }\href@noop {} {\bibfield  {journal} {\bibinfo
  {journal} {Soft Matter}\ }\textbf {\bibinfo {volume} {10}},\ \bibinfo {pages}
  {48} (\bibinfo {year} {2014})}\BibitemShut {NoStop}%
\bibitem [{\citenamefont {Hsiao}\ and\ \citenamefont
  {Pradeep}(2019)}]{hsiao2019experimental}%
  \BibitemOpen
  \bibfield  {author} {\bibinfo {author} {\bibfnamefont {L.~C.}\ \bibnamefont
  {Hsiao}}\ and\ \bibinfo {author} {\bibfnamefont {S.}~\bibnamefont
  {Pradeep}},\ }\href@noop {} {\bibfield  {journal} {\bibinfo  {journal}
  {Current Opinion in Colloid \& Interface Science}\ } (\bibinfo {year}
  {2019})}\BibitemShut {NoStop}%
\bibitem [{\citenamefont {Alonso-Marroquin}(2008)}]{alonso2008spheropolygons}%
  \BibitemOpen
  \bibfield  {author} {\bibinfo {author} {\bibfnamefont {F.}~\bibnamefont
  {Alonso-Marroquin}},\ }\href@noop {} {\bibfield  {journal} {\bibinfo
  {journal} {EPL (Europhysics Letters)}\ }\textbf {\bibinfo {volume} {83}},\
  \bibinfo {pages} {14001} (\bibinfo {year} {2008})}\BibitemShut {NoStop}%
\bibitem [{\citenamefont {Papanikolaou}\ \emph {et~al.}(2013)\citenamefont
  {Papanikolaou}, \citenamefont {O'Hern},\ and\ \citenamefont
  {Shattuck}}]{papanikolaou2013}%
  \BibitemOpen
  \bibfield  {author} {\bibinfo {author} {\bibfnamefont {S.}~\bibnamefont
  {Papanikolaou}}, \bibinfo {author} {\bibfnamefont {C.~S.}\ \bibnamefont
  {O'Hern}}, \ and\ \bibinfo {author} {\bibfnamefont {M.~D.}\ \bibnamefont
  {Shattuck}},\ }\href {\doibase 10.1103/PhysRevLett.110.198002} {\bibfield
  {journal} {\bibinfo  {journal} {Phys. Rev. Lett.}\ }\textbf {\bibinfo
  {volume} {110}},\ \bibinfo {pages} {198002} (\bibinfo {year}
  {2013})}\BibitemShut {NoStop}%
\bibitem [{\citenamefont {Donev}\ \emph {et~al.}(2007)\citenamefont {Donev},
  \citenamefont {Connelly}, \citenamefont {Stillinger},\ and\ \citenamefont
  {Torquato}}]{donev2007}%
  \BibitemOpen
  \bibfield  {author} {\bibinfo {author} {\bibfnamefont {A.}~\bibnamefont
  {Donev}}, \bibinfo {author} {\bibfnamefont {R.}~\bibnamefont {Connelly}},
  \bibinfo {author} {\bibfnamefont {F.~H.}\ \bibnamefont {Stillinger}}, \ and\
  \bibinfo {author} {\bibfnamefont {S.}~\bibnamefont {Torquato}},\ }\href
  {\doibase 10.1103/PhysRevE.75.051304} {\bibfield  {journal} {\bibinfo
  {journal} {Phys. Rev. E}\ }\textbf {\bibinfo {volume} {75}},\ \bibinfo
  {pages} {051304} (\bibinfo {year} {2007})}\BibitemShut {NoStop}%
\bibitem [{\citenamefont {Williams}\ and\ \citenamefont
  {Philipse}(2003)}]{williams2003}%
  \BibitemOpen
  \bibfield  {author} {\bibinfo {author} {\bibfnamefont {S.~R.}\ \bibnamefont
  {Williams}}\ and\ \bibinfo {author} {\bibfnamefont {A.~P.}\ \bibnamefont
  {Philipse}},\ }\href {\doibase 10.1103/PhysRevE.67.051301} {\bibfield
  {journal} {\bibinfo  {journal} {Phys. Rev. E}\ }\textbf {\bibinfo {volume}
  {67}},\ \bibinfo {pages} {051301} (\bibinfo {year} {2003})}\BibitemShut
  {NoStop}%
\bibitem [{\citenamefont {Blouwolff}\ and\ \citenamefont
  {Fraden}(2006)}]{blouwolff2006}%
  \BibitemOpen
  \bibfield  {author} {\bibinfo {author} {\bibfnamefont {J.}~\bibnamefont
  {Blouwolff}}\ and\ \bibinfo {author} {\bibfnamefont {S.}~\bibnamefont
  {Fraden}},\ }\href@noop {} {\bibfield  {journal} {\bibinfo  {journal} {EPL
  (Europhysics Letters)}\ }\textbf {\bibinfo {volume} {76}},\ \bibinfo {pages}
  {1095} (\bibinfo {year} {2006})}\BibitemShut {NoStop}%
\bibitem [{\citenamefont {VanderWerf}\ \emph {et~al.}(2018)\citenamefont
  {VanderWerf}, \citenamefont {Jin}, \citenamefont {Shattuck},\ and\
  \citenamefont {O'Hern}}]{werf2018}%
  \BibitemOpen
  \bibfield  {author} {\bibinfo {author} {\bibfnamefont {K.}~\bibnamefont
  {VanderWerf}}, \bibinfo {author} {\bibfnamefont {W.}~\bibnamefont {Jin}},
  \bibinfo {author} {\bibfnamefont {M.~D.}\ \bibnamefont {Shattuck}}, \ and\
  \bibinfo {author} {\bibfnamefont {C.~S.}\ \bibnamefont {O'Hern}},\ }\href
  {\doibase 10.1103/PhysRevE.97.012909} {\bibfield  {journal} {\bibinfo
  {journal} {Phys. Rev. E}\ }\textbf {\bibinfo {volume} {97}},\ \bibinfo
  {pages} {012909} (\bibinfo {year} {2018})}\BibitemShut {NoStop}%
\bibitem [{\citenamefont {Brito}\ \emph {et~al.}(2018)\citenamefont {Brito},
  \citenamefont {Ikeda}, \citenamefont {Urbani}, \citenamefont {Wyart},\ and\
  \citenamefont {Zamponi}}]{brito2018}%
  \BibitemOpen
  \bibfield  {author} {\bibinfo {author} {\bibfnamefont {C.}~\bibnamefont
  {Brito}}, \bibinfo {author} {\bibfnamefont {H.}~\bibnamefont {Ikeda}},
  \bibinfo {author} {\bibfnamefont {P.}~\bibnamefont {Urbani}}, \bibinfo
  {author} {\bibfnamefont {M.}~\bibnamefont {Wyart}}, \ and\ \bibinfo {author}
  {\bibfnamefont {F.}~\bibnamefont {Zamponi}},\ }\href@noop {} {\bibfield
  {journal} {\bibinfo  {journal} {Proceedings of the National Academy of
  Sciences}\ }\textbf {\bibinfo {volume} {115}},\ \bibinfo {pages} {11736}
  (\bibinfo {year} {2018})}\BibitemShut {NoStop}%
\bibitem [{\citenamefont {Ikeda}\ \emph {et~al.}(2019)\citenamefont {Ikeda},
  \citenamefont {Urbani},\ and\ \citenamefont {Zamponi}}]{ikeda2019mean}%
  \BibitemOpen
  \bibfield  {author} {\bibinfo {author} {\bibfnamefont {H.}~\bibnamefont
  {Ikeda}}, \bibinfo {author} {\bibfnamefont {P.}~\bibnamefont {Urbani}}, \
  and\ \bibinfo {author} {\bibfnamefont {F.}~\bibnamefont {Zamponi}},\
  }\href@noop {} {\bibfield  {journal} {\bibinfo  {journal} {Journal of Physics
  A: Mathematical and Theoretical}\ }\textbf {\bibinfo {volume} {52}},\
  \bibinfo {pages} {344001} (\bibinfo {year} {2019})}\BibitemShut {NoStop}%
\bibitem [{\citenamefont {Ikeda}\ \emph {et~al.}(2020)\citenamefont {Ikeda},
  \citenamefont {Brito},\ and\ \citenamefont {Wyart}}]{ikeda2020}%
  \BibitemOpen
  \bibfield  {author} {\bibinfo {author} {\bibfnamefont {H.}~\bibnamefont
  {Ikeda}}, \bibinfo {author} {\bibfnamefont {C.}~\bibnamefont {Brito}}, \ and\
  \bibinfo {author} {\bibfnamefont {M.}~\bibnamefont {Wyart}},\ }\href@noop {}
  {\bibfield  {journal} {\bibinfo  {journal} {Journal of Statistical Mechanics:
  Theory and Experiment}\ }\textbf {\bibinfo {volume} {2020}},\ \bibinfo
  {pages} {033302} (\bibinfo {year} {2020})}\BibitemShut {NoStop}%
\bibitem [{Note1()}]{Note1}%
  \BibitemOpen
  \bibinfo {note} {$F$ is not a periodic function of $u_i$ and $u_j$. But, this
  is not a big problem because such symmetries do not affect the local
  stability arguments discussed here.}\BibitemShut {Stop}%
\bibitem [{\citenamefont {Bitzek}\ \emph {et~al.}(2006)\citenamefont {Bitzek},
  \citenamefont {Koskinen}, \citenamefont {G\"ahler}, \citenamefont {Moseler},\
  and\ \citenamefont {Gumbsch}}]{fire2006}%
  \BibitemOpen
  \bibfield  {author} {\bibinfo {author} {\bibfnamefont {E.}~\bibnamefont
  {Bitzek}}, \bibinfo {author} {\bibfnamefont {P.}~\bibnamefont {Koskinen}},
  \bibinfo {author} {\bibfnamefont {F.}~\bibnamefont {G\"ahler}}, \bibinfo
  {author} {\bibfnamefont {M.}~\bibnamefont {Moseler}}, \ and\ \bibinfo
  {author} {\bibfnamefont {P.}~\bibnamefont {Gumbsch}},\ }\href {\doibase
  10.1103/PhysRevLett.97.170201} {\bibfield  {journal} {\bibinfo  {journal}
  {Phys. Rev. Lett.}\ }\textbf {\bibinfo {volume} {97}},\ \bibinfo {pages}
  {170201} (\bibinfo {year} {2006})}\BibitemShut {NoStop}%
\bibitem [{\citenamefont {Xiong}\ \emph {et~al.}(2019)\citenamefont {Xiong},
  \citenamefont {Wang}, \citenamefont {Clark}, \citenamefont {Bertrand},
  \citenamefont {Ouellette}, \citenamefont {Shattuck},\ and\ \citenamefont
  {O’Hern}}]{xiong2019}%
  \BibitemOpen
  \bibfield  {author} {\bibinfo {author} {\bibfnamefont {F.}~\bibnamefont
  {Xiong}}, \bibinfo {author} {\bibfnamefont {P.}~\bibnamefont {Wang}},
  \bibinfo {author} {\bibfnamefont {A.~H.}\ \bibnamefont {Clark}}, \bibinfo
  {author} {\bibfnamefont {T.}~\bibnamefont {Bertrand}}, \bibinfo {author}
  {\bibfnamefont {N.~T.}\ \bibnamefont {Ouellette}}, \bibinfo {author}
  {\bibfnamefont {M.~D.}\ \bibnamefont {Shattuck}}, \ and\ \bibinfo {author}
  {\bibfnamefont {C.~S.}\ \bibnamefont {O’Hern}},\ }\href@noop {} {\bibfield
  {journal} {\bibinfo  {journal} {Granular Matter}\ }\textbf {\bibinfo {volume}
  {21}},\ \bibinfo {pages} {109} (\bibinfo {year} {2019})}\BibitemShut
  {NoStop}%
\bibitem [{Note2()}]{Note2}%
  \BibitemOpen
  \bibinfo {note} {For large $n$, $z_J$ has a minimum at an intermediate value
  of $\mu $. This counter-intuitive behavior could be a signature of the
  breakdown of our perturbative approach. The position of the minimum shifts to
  higher $\mu $ upon increasing of $n$, and the result of the Coulomb friction
  law is correctly reproduced in the limit of $n\to \infty $.}\BibitemShut
  {Stop}%
\bibitem [{\citenamefont {Song}\ \emph {et~al.}(2008)\citenamefont {Song},
  \citenamefont {Wang},\ and\ \citenamefont {Makse}}]{song2008phase}%
  \BibitemOpen
  \bibfield  {author} {\bibinfo {author} {\bibfnamefont {C.}~\bibnamefont
  {Song}}, \bibinfo {author} {\bibfnamefont {P.}~\bibnamefont {Wang}}, \ and\
  \bibinfo {author} {\bibfnamefont {H.~A.}\ \bibnamefont {Makse}},\ }\href@noop
  {} {\bibfield  {journal} {\bibinfo  {journal} {Nature}\ }\textbf {\bibinfo
  {volume} {453}},\ \bibinfo {pages} {629} (\bibinfo {year}
  {2008})}\BibitemShut {NoStop}%
\bibitem [{\citenamefont {Shen}\ \emph {et~al.}(2014)\citenamefont {Shen},
  \citenamefont {Papanikolaou}, \citenamefont {O'Hern},\ and\ \citenamefont
  {Shattuck}}]{shen2014}%
  \BibitemOpen
  \bibfield  {author} {\bibinfo {author} {\bibfnamefont {T.}~\bibnamefont
  {Shen}}, \bibinfo {author} {\bibfnamefont {S.}~\bibnamefont {Papanikolaou}},
  \bibinfo {author} {\bibfnamefont {C.~S.}\ \bibnamefont {O'Hern}}, \ and\
  \bibinfo {author} {\bibfnamefont {M.~D.}\ \bibnamefont {Shattuck}},\ }\href
  {\doibase 10.1103/PhysRevLett.113.128302} {\bibfield  {journal} {\bibinfo
  {journal} {Phys. Rev. Lett.}\ }\textbf {\bibinfo {volume} {113}},\ \bibinfo
  {pages} {128302} (\bibinfo {year} {2014})}\BibitemShut {NoStop}%
\end{thebibliography}%
\end{document}